\documentclass[twocolumn,aps,prl,superscriptaddress,nofootinbib,floatfix,longbibliography]{revtex4-2}

\usepackage{amsmath,amssymb}
\usepackage{graphicx}
\usepackage{xcolor}
\usepackage{booktabs}
\usepackage{orcidlink}
\usepackage{hyperref}

\hypersetup{
  colorlinks=true,
  linkcolor=blue!50!black,
  citecolor=blue!50!black,
  urlcolor=blue!50!black
}

\newcommand{\DS}{\Delta S_A}
\newcommand{\rhoAsym}{\rho_A^{\rm sym}}
\newcommand{\fcross}{f_{\rm cross}}

\begin{document}

\title{Selective Quantum Mpemba Dynamics in Chaotic Spin Chains}

\author{Ri-Hua Zheng\orcidlink{0000-0002-1944-1573}}
\author{Yang Xiao\orcidlink{0009-0000-7074-8805}}
\affiliation{Fujian Key Laboratory of Quantum Information and Quantum Optics,
College of Physics and Information Engineering, Fuzhou University, Fuzhou,
Fujian 350108, China}
\author{Yu Wang\orcidlink{0000-0003-4399-946X}}
\affiliation{School of Physics, Hangzhou Normal University, Hangzhou 311121,
China}
\author{Ye-Hong Chen\orcidlink{0000-0002-7308-2823}}\thanks{yehong.chen@fzu.edu.cn}
\affiliation{Fujian Key Laboratory of Quantum Information and Quantum Optics,
College of Physics and Information Engineering, Fuzhou University, Fuzhou,
Fujian 350108, China}
\affiliation{Quantum Information Physics Theory Research Team, Center for
Quantum Computing, RIKEN, Wako-shi, Saitama 351-0198, Japan}
\author{Yan Xia\orcidlink{0000-0002-4539-298X}}\thanks{xia-208@163.com}
\affiliation{Fujian Key Laboratory of Quantum Information and Quantum Optics,
College of Physics and Information Engineering, Fuzhou University, Fuzhou,
Fujian 350108, China}
\date[]{}

\begin{abstract}
In a symmetry-restoration quantum Mpemba effect, an initial state with stronger
local symmetry breaking can lose that memory faster than a state that starts
closer to the symmetric manifold.  We study how this local ordering reversal
depends on the spatial profile of a longitudinal field in a clean
$U(1)$-conserving spin chain, comparing spectral level statistics with crossings of the entanglement
asymmetry for the same Hamiltonians.  We find that Gaussian orthogonal ensemble
(GOE)-like level statistics alone do not determine whether Mpemba crossings
occur.  Across field textures, GOE-like spectra can occur with or without
entanglement-asymmetry crossings at the same field strength.  In a near-staggered detuned field, the total charge-sector
coherence reverses its ordering while the entanglement asymmetry retains its
initial order.  This contrast links the crossing response to the distribution
of local charge-sector coherence and its entropic weighting in the reduced
density matrix.
\end{abstract}

\maketitle

\begin{figure}[t]
\centering
\includegraphics[width=\columnwidth]{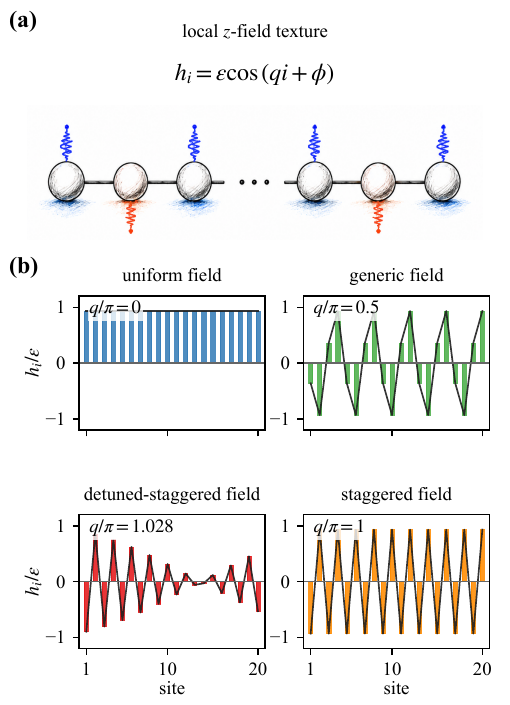}
\caption{
Model and field textures.  (a) Schematic of the structured longitudinal
$z$-field $h_i$ applied to a clean $U(1)$-conserving XXZ chain.  The upward
blue and downward red arrows indicate the sign and spatial pattern of the local
field bias; $q$ sets its wave number, $\epsilon$ its amplitude, and $\phi$ a
fixed offset relative to the open-chain boundaries.  Because the perturbation
is longitudinal, it preserves total $S^z$.  (b) Representative $N=20$ field
textures used throughout this work: uniform field ($q=0$, same sign and
magnitude on all sites), generic field ($q/\pi=0.5$, non-staggered spatial
modulation), detuned-staggered field ($q/\pi=1.028$,
detuned from exact staggering), and staggered field ($q=\pi$, alternating
site by site).
}
\label{fig:model}
\end{figure}

Relaxation is often viewed as a monotone loss of memory, yet the Mpemba effect
shows that a seemingly more distant state can approach stationarity faster than
a milder one.  First posed for water freezing \cite{Mpemba1969,Jeng2006}, it has
since been observed and modeled in granular fluids \cite{Lasanta2017}, Markov
processes and anomalous-relaxation models \cite{LuRaz2017,Klich2019}, spin
glasses \cite{BaityJesi2019}, thermal protocols \cite{GalRaz2020}, and colloidal
experiments \cite{Kumar2020}.  In quantum many-body settings, the analogue is
whether relaxation can reorder local symmetry breaking, so that an initially
more asymmetric state later retains less asymmetry.

For symmetry restoration, that ordering is naturally measured by the
entanglement asymmetry.  After a unitary quench in a system with a globally
conserved charge, a subsystem can restore the symmetry broken by the initial
state; entanglement asymmetry tracks this restoration through charge-sector
coherence in the reduced density matrix
\cite{Ares2023EntAsym,Ares2023Lack}.  A quantum Mpemba crossing occurs when the
state with larger initial asymmetry later becomes less asymmetric than one that
began closer to the symmetric manifold
\cite{Murciano2024XY,Rylands2024Microscopic,AresReview2025}.  Related quantum
Mpemba effects have been formulated in open systems through Liouvillian decay
modes and reservoir-controlled relaxation
\cite{Carollo2021,Chatterjee2023QuantumDot,Chatterjee2024Multiple,Nava2024}, with
extensions to thermodynamic, non-Markovian, and experimental settings
\cite{Moroder2024,Strachan2025,AharonyShapira2024,Zhang2025Strong}.

Chaotic thermalization is a candidate mechanism for such an ordering
reversal.  Integrable spectra cluster whereas chaotic spectra repel and approach
Gaussian orthogonal ensemble (GOE) statistics
\cite{BerryTabor1977,Bohigas1984,Atas2013}, and the eigenstate thermalization
hypothesis connects nonintegrable eigenstates to thermal local physics
\cite{Deutsch1991,Srednicki1994,Rigol2008,DAlessio2016}.  Symmetry-restoration
crossings have been studied in integrable, long-range, two-dimensional,
mixed-state, measurement-induced, and non-symmetric settings
\cite{Rylands2024Heisenberg,Yamashika2024TwoD,Chalas2024Multiple,
AresVitale2025Mixed,DiGiulio2025,Bhore2025NoSym,Yu2025Tuning,
AresRylands2025,YamashikaAres2025,Joshi2024Obs,YuReview2025}, while
integrability breaking, resonances, and chaotic systems with conservation laws
provide a more direct route to the role of chaos
\cite{Kinoshita2006,Kaufman2016,CalabreseCardy2005,
McRoberts2026,YamashikaHamazaki2026,Muller2026}.

In disordered many-body systems, level statistics are widely used alongside
transport and local-memory diagnostics to study ergodicity and localization
\cite{OganesyanHuse2007,PalHuse2010,Luitz2015,NandkishoreHuse2015,Abanin2019}.
For the quantum Mpemba effect, the dynamical question concerns the relative
relaxation of different initial states, as revealed by crossings of their
subsystem entanglement asymmetries.  We ask whether spectral chaos determines
this ordering reversal within a single clean Hamiltonian family.

Here we vary the longitudinal-field texture in a clean $U(1)$-conserving XXZ
chain \cite{Bethe1931} and compare the two diagnostics for each Hamiltonian.
The generic, staggered, and near-staggered detuned fields have similar level
statistics, approaching the GOE reference at the larger sizes studied.  Yet
the first two fields produce
entanglement-asymmetry crossings, whereas the detuned field preserves the
initial ordering.  We further use this detuned
control to test a more local diagnostic: the total off-block-diagonal charge
weight of the reduced density matrix.  In this control, the total
charge-coherence ordering inverts while the entanglement-asymmetry ordering
persists.  Together, these results show that neither spectral chaos nor the
total charge-sector coherence determines the Mpemba crossing.  Resolving the
coherence by charge sector reveals competing contributions with different
entropic weights.

We use an open spin-$1/2$ antiferromagnetic XXZ chain \cite{Bethe1931}:
\begin{equation}
H_0=\sum_{i=1}^{N-1}
J
\left[
\frac{1}{2}(S_i^+S_{i+1}^-+S_i^-S_{i+1}^+)
+\Delta S_i^zS_{i+1}^z
\right],
\end{equation}
perturbed by a longitudinal field texture
\begin{equation}
H=H_0+\sum_{i=1}^N h_i S_i^z,\qquad
h_i=\epsilon\cos(qi+\phi).
\label{eq:model}
\end{equation}
We work at the antiferromagnetic isotropic Heisenberg point, setting the
exchange scale $J=1$ and the exchange anisotropy $\Delta=1$.
Here $N$ is the number of spins, the site index is $i=1,\ldots,N$, and
$S_i^\alpha$ are spin-$1/2$ operators with
$S_i^\pm=S_i^x\pm {\rm i}S_i^y$.  The exchange $J$ sets the energy and time
units, $\epsilon$ is the longitudinal-field strength, $q$ is the field wave
number, and $\phi$ is a fixed phase offset.  The perturbation preserves total
charge,
$[H,\sum_iS_i^z]=0$, while breaking integrability for generic field profiles.
To isolate the effect of spatial texture, we hold the field strength and phase
fixed across all textures and sizes, choosing $\epsilon=0.6$ and $\phi=0.37$.
The phase sets a
nonspecial origin of the cosine relative to the open-chain boundaries and
subsystem cuts.
Figure~\ref{fig:model}
summarizes the model and defines the four representative field textures:
uniform fields ($q=0$), for which all sites experience the same longitudinal
field; generic fields ($q/\pi=0.5$), which provide a non-staggered spatial
modulation; staggered fields ($q=\pi$), where the field alternates from site to
site; and detuned-staggered fields ($q/\pi=1.028$ at every size), which remain close to the staggered pattern
but depart from its exact two-site periodicity.  The detuned-staggered fields provide
clean no-crossing controls near the staggered limit.  As local $S^z$ biases,
these longitudinal textures reshape how charge-sector phases and local charge
motion develop under the charge-conserving exchange dynamics
\cite{Ares2023EntAsym,Ares2023Lack,Rylands2024Heisenberg}.

The local symmetry-restoration observable is the entanglement asymmetry.  For a
contiguous subsystem $A$,
\begin{align}
\DS(t)&=S[\rhoAsym(t)]-S[\rho_A(t)], \\
\rhoAsym(t)&=\sum_n P_n\rho_A(t)P_n ,
\label{eq:asymmetry}
\end{align}
where $S(\rho)=-{\rm Tr}\rho\ln\rho$ is the von Neumann entropy.  The subsystem
charge is $Q_A=\sum_{i\in A}S_i^z$, and $P_n$ projects onto the eigenspace with
subsystem charge $Q_A=n-L_A/2$, where $n$ is the number of up spins in $A$.
The symmetrization removes the off-diagonal blocks
between different $Q_A$ sectors.  Thus $\DS(t)$ measures, in entropic units, how
much local $U(1)$-breaking charge-sector coherence remains in $\rho_A(t)$.
Further details are given in the Supplemental Material, Sec.~S1
\cite{SupplementalMaterial}.

Initial states are tilted ferromagnetic product states
\begin{equation}
|\psi(\theta)\rangle=
\bigotimes_i
\left(
\cos\frac{\theta}{2}|\uparrow\rangle_i+
\sin\frac{\theta}{2}|\downarrow\rangle_i
\right),
\end{equation}
where $|\uparrow\rangle_i$ and $|\downarrow\rangle_i$ are the local
$S_i^z=\pm1/2$ eigenstates.
To sample initial symmetry breaking, we use the same fixed, equally spaced
ladder for every field texture,
$\theta/\pi=0.28,0.32,0.36,0.40,0.44,0.48$.  Increasing $\theta$ rotates the
product spin from the conserved $z$ direction toward the transverse plane,
giving well separated initial values of $\DS(0)$ while keeping the state family
fixed.  For a pair satisfying
$\Delta S_{A,a}(0)>\Delta S_{A,b}(0)$, a raw QME crossing occurs when
$\Delta S_{A,a}(t)<\Delta S_{A,b}(t)$ for a sustained time interval.  We summarize the response by
\begin{equation}
\fcross=
\frac{\text{number of sustained crossing pairs}}
{\text{number of pairs with unequal initial }\DS(0)}.
\end{equation}
For the data below, a crossing must persist for at least one unit of the
dimensionless time $Jt$ with $J=1$.  We
average $\DS(t)$ over all contiguous subsystems of the same length
$L_A$, the size of subsystem $A$.  The
subsystem is chosen to be roughly one quarter of the chain: large enough to
resolve charge-sector coherence, but still local enough to probe symmetry
restoration inside a finite isolated system.  For each fixed $N$, all four field
textures are compared at the same $L_A$.  The precise raw-crossing protocol and
subsystem-size choices are described in the Supplemental Material, Sec.~S2
\cite{SupplementalMaterial}.

\begin{figure}[t]
\centering
\includegraphics[width=\linewidth]{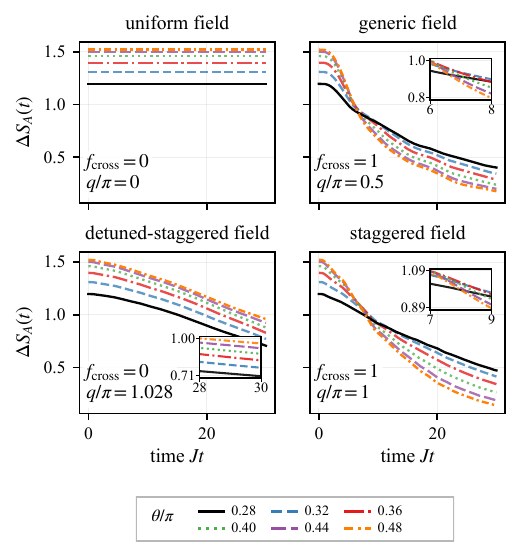}
\caption{
Representative $N=20$ entanglement-asymmetry dynamics.  Lines correspond to
tilted ferromagnetic product states with the six initial angles indicated by
the discrete color legend; all four panels use the exchange-scale field
strength $\epsilon=0.6$.  The generic and staggered fields have $\fcross=1$,
while the detuned-staggered control has $\fcross=0$.  The insets magnify the
crossing windows in the generic and staggered panels and the late-time window
where the detuned-staggered curves are closest but remain ordered.
}
\label{fig:n20curves}
\end{figure}

Figure~\ref{fig:n20curves} shows the largest-size dynamics.  In the generic and
staggered fields, every pair with unequal initial asymmetry crosses.  The detuned-staggered
asymmetry curves decay while retaining their initial ordering over the simulated
time window.

To compare this local response with spectral chaos, we compute level statistics
within fixed total-$S^z$ sectors, as required by the conserved charge of
Eq.~\eqref{eq:model}.  The spectral statistic is the adjacent-gap ratio
\begin{equation}
r_j=
\frac{\min(\delta_j,\delta_{j+1})}
{\max(\delta_j,\delta_{j+1})},
\qquad
\delta_j=E_{j+1}-E_j ,
\end{equation}
where $E_j$ are consecutive sorted eigenenergies in the chosen symmetry sector.
Gap averaging gives the reference values
$\langle r\rangle_{\rm Poisson}\simeq0.386$ and
$\langle r\rangle_{\rm GOE}\simeq0.531$.  Details of the mid-spectrum
calculation, including total-spin and reflection resolution for uniform fields
and combined reflection--spin-flip resolution for staggered fields, are
given in the
Supplemental Material, Sec.~S3 \cite{SupplementalMaterial}.

\begin{figure}[t]
\centering
\includegraphics[width=\linewidth]{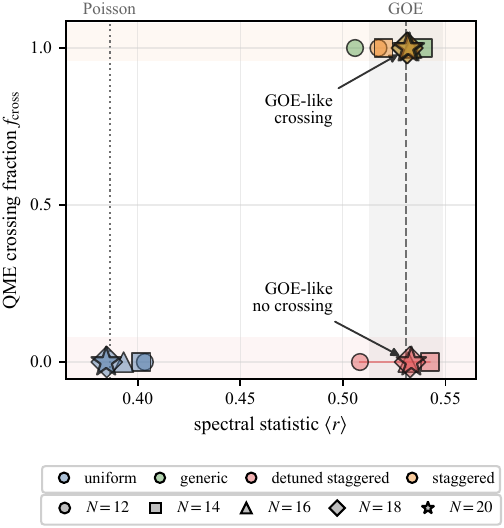}
\caption{
Selective relation between spectral chaos and raw quantum-Mpemba crossings.
Each point combines $\langle r\rangle$ and $\fcross$ for the same Hamiltonian at
fixed $\epsilon=0.6$.  Colors denote field texture and markers denote system size;
all four textures are included at $N=12,14,16,18,20$.  Generic and staggered
fields have $\fcross=1$, while detuned-staggered fields retain $\fcross=0$
at similar GOE-like gap ratios.  Shading serves as a visual guide.
}
\label{fig:scatter}
\end{figure}

Figure~\ref{fig:scatter} compares the spectral and QME responses under the
same field parameters.  The crossing response separates by field texture:
generic and staggered textures form the crossing class, while uniform and
detuned-staggered textures have $\fcross=0$.  The two dynamical classes overlap
in spectral statistics.  At the larger sizes studied, the generic, staggered,
and detuned-staggered fields have nearly the same gap ratio while retaining
opposite crossing responses; at $N=20$, all three are tightly clustered around
the GOE reference.  The uniform field provides a Poisson-like no-crossing
reference.  Thus GOE-like level statistics can accompany opposite Mpemba
crossing responses within this clean $U(1)$-conserving Hamiltonian family at
finite sizes.  Level repulsion characterizes the Hamiltonian spectrum, whereas
a Mpemba crossing is a relational property of two initial-state trajectories.
The gap ratio uses only eigenenergies in a fixed charge sector, while subsystem
dynamics also depends on initial-state overlaps with energy eigenstates and on
their structure across the subsystem boundary.  Similar spectral statistics
therefore need not imply similar relative relaxation.  For entanglement
asymmetry, the relevant local information lies in the charge-sector structure
of $\rho_A(t)$.  Coherence between distinct subsystem charge sectors breaks
local symmetry, and $\DS$ measures this asymmetry through the entropy difference
between $\rhoAsym$ and $\rho_A$.  The asymmetries of two initial states evolving
under the same Hamiltonian can both decrease; a Mpemba crossing occurs only when
their relative decay reverses their initial ordering.  The spectral comparison
thus directs attention to how local charge coherence evolves for different
initial states.

We next test whether inversion of the total charge-sector coherence accounts
for the entanglement-asymmetry crossings, using the Frobenius measure
\begin{equation}
C_A(t)=
\left\|\rho_A(t)-\sum_nP_n\rho_A(t)P_n\right\|_F^2 ,
\end{equation}
which measures the total off-block-diagonal charge weight.  The distinction
from entanglement asymmetry is exact.  Writing
$\mathcal D_A(\rho_A)=\rhoAsym$, one has
\begin{equation}
\DS=D\!\left(\rho_A\middle\|\rhoAsym\right),
\label{eq:relative_entropy}
\end{equation}
where $D(\rho\|\sigma)=\mathrm{Tr}[\rho(\ln\rho-\ln\sigma)]$ is the quantum
relative entropy.  The norm $C_A$ assigns unit weight to each squared
off-block matrix element.  The weak-coherence expansion of $\DS$ makes the
eigenvalue-dependent weights explicit; its form is given in Supplemental Material,
Sec.~S4 \cite{SupplementalMaterial}.  For either $X=C_A$ or $X=\DS$, we retain the
depth and duration of the strongest ordering inversion through
\begin{equation}
R_X^*=\max_{a,b:\,X_a(0)>X_b(0)}
\frac{1}{T}\int_0^T dt\,\bigl[X_b(t)-X_a(t)\bigr]_+,
\label{eq:inversion_area}
\end{equation}
where $[x]_+=\max(x,0)$, $T=30$, and the maximum runs over all distinct
state pairs satisfying $X_a(0)>X_b(0)$.  Thus $R_X^*=0$ means that no pair reverses its ordering,
whereas a nonzero value retains both the depth and duration of the strongest
inversion.

\begin{figure}[b]
\centering
\includegraphics[width=\linewidth]{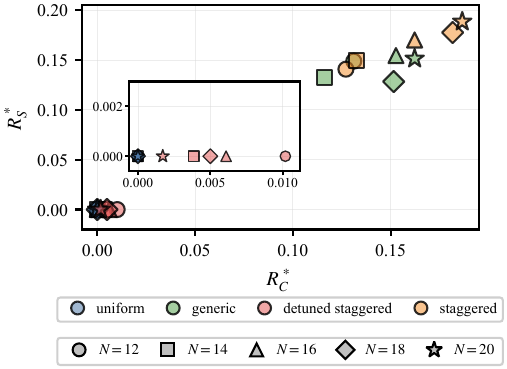}
\caption{Local charge-coherence inversion versus entropic ordering.  Each point
combines $R_C^*$ and $R_S^*$ for the same Hamiltonian at fixed $\epsilon=0.6$;
colors denote texture and markers denote size.  Uniform points lie at the
origin, generic and staggered points are nonzero on both axes, and
detuned-staggered points have $R_C^*>0$ but $R_S^*=0$ at every displayed size.
The inset resolves these points near the origin.  Total
charge-coherence inversion is therefore not sufficient for an
entanglement-asymmetry Mpemba crossing.}
\label{fig:coherence_area}
\end{figure}

Figure~\ref{fig:coherence_area} compares the two inversion measures across
different system sizes.  Uniform fields invert in neither diagnostic, and generic and
staggered fields invert in both.  Detuned-staggered fields retain $R_S^*=0$
while $R_C^*>0$ at every displayed size.  Its strength varies nonmonotonically, but
the separation between the two orderings persists.  The unweighted Frobenius
norm and the relative entropy in Eq.~\eqref{eq:relative_entropy} therefore
order the same evolving states differently.
Definitions, crossing
fractions, and representative raw curves are given in the Supplemental
Material, Sec.~S4 and Fig.~S1 \cite{SupplementalMaterial}.

To identify the origin of this mismatch, we use the weak-coherence expansion
to resolve the competition among entropy-weighted charge-sector contributions
at $N=20$.
For the representative pair $\theta/\pi=0.36,0.48$, we subtract the initially less
asymmetric state's block contribution from the initially more asymmetric
state's contribution.  Positive differences preserve the initial ordering,
whereas negative differences favor inversion.
At $t=10$, high-charge edge blocks favor inversion in both fields.
Their negative contribution outweighs the positive central-sector terms in
the generic field.  In the detuned-staggered field, stronger positive terms
instead preserve the entropy-weighted ordering, even though the unweighted
coherence has inverted.  The quadratic diagnostic reproduces this sign
contrast in the exact entanglement asymmetry; its quantitative accuracy and
time dependence are examined in Supplemental Material, Sec.~S4 and Fig.~S2
\cite{SupplementalMaterial}.

We have shown that field texture organizes the raw QME crossing response in
this clean $U(1)$-conserving spin chain.  Changing only the local $z$-field profile separates
three levels of information: global level statistics, the total local
charge-sector coherence, and the entropy-weighted ordering encoded in
entanglement asymmetry.  In the detuned-staggered field, a GOE-like spectrum
coexists with inversion of the total local coherence and persistent
entanglement-asymmetry ordering across the different sizes studied.  The $N=20$
quadratic block analysis resolves competing charge-sector contributions and
connects their eigenvalue-dependent weights to the different entropy orderings
in the generic and detuned-staggered fields.  Thus the relative loss of local
symmetry-breaking memory depends on information that neither spectral level
repulsion nor a single coherence norm retains.

\begin{acknowledgments}
Y.W. was supported by the National Natural Science Foundation of China (Grant
No. 12404404).
Y.-H.C. was supported by the National Natural Science Foundation of China
(Grants No. 12304390 and No. 12574386), the National Postdoctoral Overseas
Talent Recruitment Program of China, the Fujian 100 Talents Program, and the
Fujian Minjiang Scholar Program. Y.X. was supported by the National Natural
Science Foundation of China under Grant No. 62471143, the Key Program of
National Natural Science Foundation of Fujian Province under Grant
No. 2024J02008.
\end{acknowledgments}

\bibliography{references}

\end{document}

\title{Supplemental Material for\\
``Selective Quantum Mpemba Dynamics in Chaotic Spin Chains''}

\author{Ri-Hua Zheng\orcidlink{0000-0002-1944-1573}}
\author{Yang Xiao\orcidlink{0009-0000-7074-8805}}
\affiliation{Fujian Key Laboratory of Quantum Information and Quantum Optics,
College of Physics and Information Engineering, Fuzhou University, Fuzhou,
Fujian 350108, China}
\author{Yu Wang\orcidlink{0000-0003-4399-946X}}
\affiliation{School of Physics, Hangzhou Normal University, Hangzhou 311121,
China}
\author{Ye-Hong Chen\orcidlink{0000-0002-7308-2823}}
\affiliation{Fujian Key Laboratory of Quantum Information and Quantum Optics,
College of Physics and Information Engineering, Fuzhou University, Fuzhou,
Fujian 350108, China}
\affiliation{Quantum Information Physics Theory Research Team, Center for
Quantum Computing, RIKEN, Wako-shi, Saitama 351-0198, Japan}
\author{Yan Xia\orcidlink{0000-0002-4539-298X}}
\affiliation{Fujian Key Laboratory of Quantum Information and Quantum Optics,
College of Physics and Information Engineering, Fuzhou University, Fuzhou,
Fujian 350108, China}

\date[]{}
\maketitle

\begin{center}
\textbf{CONTENTS}
\end{center}

\noindent
\begin{tabular*}{\textwidth}{@{}p{0.84\textwidth}@{\extracolsep{\fill}}r@{}}
\hyperref[sec:model]{S1. Model and observables} & \pageref{sec:model} \\[0.7em]
\hyperref[sec:crossing]{S2. Raw crossing fraction} & \pageref{sec:crossing} \\[0.7em]
\hyperref[sec:level]{S3. Level statistics} & \pageref{sec:level} \\[0.7em]
\hyperref[sec:coherence]{S4. Charge-sector coherence and entropy weighting} & \pageref{sec:coherence} \\
\end{tabular*}

\clearpage

\section{S1. Model and observables}
\label{sec:model}

The Hamiltonian used in the main text is an open spin-$1/2$ antiferromagnetic
XXZ chain:
\begin{equation}
H_0=\sum_{i=1}^{N-1}J\left[
\frac{1}{2}\left(S_i^+S_{i+1}^-+S_i^-S_{i+1}^+\right)
+\Delta S_i^zS_{i+1}^z
\right],
\end{equation}
perturbed by a longitudinal field texture
\begin{equation}
H=H_0+\sum_{i=1}^{N} h_i S_i^z,\qquad
h_i=\epsilon\cos(qi+\phi),
\label{eq:supp_model}
\end{equation}
We work at the antiferromagnetic isotropic Heisenberg point, with exchange scale
$J=1$ and exchange anisotropy $\Delta=1$.
Here $N$ is the number of spins, $i=1,\ldots,N$ is the site index,
$S_i^\alpha$ are spin-$1/2$ operators, $\epsilon$ is the field strength, $q$ is
the field wave number, and $\phi$ is the phase offset.  We prescribe the same
phase and strength for all field textures and sizes to isolate their spatial
structure, choosing $\phi=0.37$ and $\epsilon=0.6$.
The wave numbers are $q/\pi=0,0.5,1.028,1$ for the uniform, generic,
detuned-staggered, and staggered fields, respectively.  The phase fixes a nonspecial origin of the
cosine relative to the open-chain boundaries and subsystem cuts.  The field preserves the total
charge $Q=\sum_i S_i^z$, so the dynamics is unitary within each global
$U(1)$ charge sector even when the field texture breaks integrability.

For a contiguous subsystem $A$, the entanglement asymmetry is
\cite{Ares2023EntAsym,Ares2023Lack}
\begin{align}
\DS(t)&=S[\rhoAsym(t)]-S[\rho_A(t)],\\
\rhoAsym(t)&=\sum_n P_n\rho_A(t)P_n ,
\end{align}
where $S(\rho)=-{\rm Tr}\rho\ln\rho$ is the von Neumann entropy,
$Q_A=\sum_{i\in A}S_i^z$ is the local charge, and $P_n$ projects the
subsystem onto the eigenspace with $n$ up spins, equivalently $Q_A=n-L_A/2$.
The operation removes off-diagonal blocks
between different $Q_A$ sectors.  The curves shown in the main text are
averaged over all contiguous subsystems of the same length $L_A$; for an open
chain these are the windows
$A=\{1,\ldots,L_A\},\{2,\ldots,L_A+1\},\ldots,\{N-L_A+1,\ldots,N\}$.

The initial product states are tilted states,
\begin{equation}
|\psi(\theta)\rangle=\bigotimes_{i=1}^{N}
\left[
\cos\frac{\theta}{2}\,|\uparrow\rangle_i
+\sin\frac{\theta}{2}\,|\downarrow\rangle_i
\right],
\end{equation}
where $|\uparrow\rangle_i$ and $|\downarrow\rangle_i$ are the local
$S_i^z=\pm1/2$ eigenstates.
For the raw-crossing scans used in Figs.~2 and 3 of the main text, we sample the
same initial-symmetry-breaking range for every field texture with the fixed,
equally spaced ladder
$\theta/\pi=0.28,0.32,0.36,0.40,0.44,0.48$.  Within this range, larger
$\theta$ gives larger initial entanglement asymmetry.  The crossing fraction
compares the same 15 pairs with unequal initial entanglement asymmetries for
every field texture.

\section{S2. Raw crossing fraction}
\label{sec:crossing}

The quantity plotted on the vertical axis of Fig.~3 is the raw crossing
fraction $\fcross$.  For a fixed Hamiltonian and subsystem size, we order each
pair of initial states by their initial entanglement asymmetry, retaining pairs
with strictly unequal values.  A pair is counted as crossed if the
initially more asymmetric state becomes less asymmetric for a sustained time
window of at least one unit of the dimensionless time $Jt$,
\begin{equation}
\Delta S_{A,{\rm high}}(t)<\Delta S_{A,{\rm low}}(t),
\end{equation}
with a numerical tolerance of $10^{-10}$.  The reported fraction is
\begin{equation}
\fcross=\frac{\hbox{number of sustained crossing pairs}}
{\hbox{number of pairs with unequal initial }\DS(0)}.
\end{equation}
All dynamical comparisons use $0\le Jt\le 30$ and 181 equally spaced time
points.  Time evolution is evaluated by applying the sparse-matrix exponential
to each initial state, with observables calculated for each subsystem position before
averaging.  The representative $N=20$ curves in Fig.~2 use $L_A=5$.
The subsystem sizes used throughout Figs.~2--4 are listed in
Table~\ref{tab:la_choices}; within each fixed $N$, all
four field textures are evaluated at the same $L_A$.

\begin{table}[h]
\caption{\label{tab:la_choices}
Subsystem sizes used for the dynamical comparisons in Figs.~2--4.}
\begin{ruledtabular}
\begin{tabular}{cc}
$N$ & $L_A$ \\
\hline
12 & 3 \\
14 & 3 \\
16 & 4 \\
18 & 5 \\
20 & 5 \\
\end{tabular}
\end{ruledtabular}
\end{table}

The reported fractions describe ordering reversals between experimentally
preparable initial states over the simulated time window.

\section{S3. Level statistics}
\label{sec:level}

Spectral statistics are computed in the half-filled global charge sector.  For
$N=20$ this sector has dimension
\begin{equation}
\binom{20}{10}=184756.
\end{equation}
From eigenvalues near the center of the many-body spectrum we form adjacent
gap ratios
\begin{equation}
r_j=\frac{\min(\delta_j,\delta_{j+1})}
{\max(\delta_j,\delta_{j+1})},\qquad
\delta_j=E_{j+1}-E_j ,
\end{equation}
and average over the retained ratios.  The reference values are
$\langle r\rangle_{\rm P}\simeq 0.386$ for Poisson statistics and
$\langle r\rangle_{\rm GOE}\simeq 0.531$ for the Gaussian orthogonal ensemble
\cite{Atas2013}.

All spectra use the parameters of Sec.~S1, including
$\epsilon=0.6$ and the size-independent detuning $q/\pi=1.028$.
We resolve the remaining symmetries within total $S^z=0$ before forming gaps.
For the uniform field, the Zeeman term vanishes in this sector, leaving the
isotropic Heisenberg Hamiltonian.  Its levels are resolved by total spin $S$
and reflection parity $R$.  Spin-flip parity $F$ is fixed by $S$ at zero
magnetization and does not replace total-spin resolution.  For the staggered
field on the even open chains considered here, reflection and spin flip each
reverse the field term, while their product $RF$ is a symmetry.  We therefore
form ratios separately in the $RF=+1$ and $RF=-1$ blocks.  The generic and
detuned-staggered fields break these reflection and spin-flip symmetries, and
their ratios are formed directly within the fixed-charge sector.

We select the $\lceil0.20D\rceil$ levels closest to the spectral trace center
$\bar E=\mathrm{Tr}(H)/D=-1/4$, where $D=\binom{N}{N/2}$.
This defines one common energy window for all symmetry blocks of a given
Hamiltonian.  After restricting each block to that window, we form adjacent
gap ratios within the block and average all retained ratios.  Equivalently,
block means are weighted by their numbers of ratios; eigenenergies from
different blocks are never pooled to form gaps.

For large sectors, shift-invert diagonalization is carried out in disjoint
spectral slices.  Each slice is checked against an independent eigenvalue
count obtained from the inertia of the shifted Hamiltonian.  Eigenvector
residuals, orthogonality, symmetry labels, and numerical resolution of
same-block gaps are checked before assembling the complete bulk window.
Dense calculations at smaller sizes and independent checks of the uniform
total-spin resolution provide cross-validation.  The uniform spectrum in
$S^z=0$ is independent of field amplitude, so the same resolved energies apply
at $\epsilon=0.6$.

\begin{table}[h]
\caption{\label{tab:n20_fig3}
Symmetry-resolved mean gap ratios underlying Fig.~3, using the trace-centered
20\% window at $\epsilon=0.6$.  The crossing fractions are $0,1,0,1$ for
the four columns, respectively, at every size.}
\begin{ruledtabular}
\begin{tabular}{ccccc}
$N$ & uniform & generic & detuned-staggered & staggered \\
\hline
12 & 0.4035 & 0.5059 & 0.5083 & 0.5174 \\
14 & 0.4017 & 0.5389 & 0.5424 & 0.5198 \\
16 & 0.3931 & 0.5336 & 0.5318 & 0.5331 \\
18 & 0.3849 & 0.5311 & 0.5328 & 0.5314 \\
20 & 0.3847 & 0.5324 & 0.5334 & 0.5316 \\
\end{tabular}
\end{ruledtabular}
\end{table}

Table~\ref{tab:n20_fig3} shows that the three nonuniform textures have nearly
identical GOE-like ratios at the larger sizes, although their crossing
fractions differ.  Smaller systems exhibit larger finite-size and window
fluctuations.  We use 20\% consistently in Fig.~3 and repeat the calculation
with nested 5\% and 10\% windows as a sensitivity check.  At $N=20$, all three
choices give the same spectral comparison (Table~\ref{tab:uniform_parity}).

\begin{table}[h]
\caption{\label{tab:uniform_parity}
$N=20$ spectral-window check.  Each entry uses the same symmetry resolution
and trace-center prescription as the main 20\% comparison.}
\begin{ruledtabular}
\begin{tabular}{lccc}
texture & 5\% & 10\% & 20\% \\
\hline
uniform & 0.3858 & 0.3854 & 0.3847 \\
generic & 0.5323 & 0.5319 & 0.5324 \\
detuned-staggered & 0.5353 & 0.5337 & 0.5334 \\
staggered & 0.5306 & 0.5326 & 0.5316 \\
\end{tabular}
\end{ruledtabular}
\end{table}

\section{S4. Charge-sector coherence and entropy weighting}
\label{sec:coherence}

The entanglement asymmetry is nonlinear in $\rho_A$ because it compares two
entropies.  To resolve the off-block-diagonal structure underlying symmetry
restoration, we also compute the Frobenius charge-sector coherence
\begin{equation}
\CA(t)=
\left\|
\rho_A(t)-\sum_n P_n\rho_A(t)P_n
\right\|_F^2 .
\label{eq:supp_coherence}
\end{equation}
This diagnostic measures the off-block-diagonal weight of $\rho_A$ in the
subsystem charge decomposition before entropy weighting.  Main-text Fig.~4
compares its ordering inversions with those of the entanglement asymmetry.

Charge dephasing, $\rhoAsym=\mathcal D_A(\rho_A)$, gives the exact identity
\begin{equation}
\DS=D\!\left(\rho_A\middle\|\rhoAsym\right),
\label{eq:supp_relative_entropy}
\end{equation}
where
$D(\rho\|\sigma)=\mathrm{Tr}[\rho(\ln\rho-\ln\sigma)]$.  Let
$\sigma=\rhoAsym$, $Y=\rho_A-\sigma$, and
$\sigma|\mu\rangle=\lambda_\mu|\mu\rangle$.  When $Y$ is perturbative on the
relevant full-rank support, the relative entropy has the expansion
\begin{equation}
\DS=
\frac{1}{2}\sum_{\mu,\nu}|Y_{\mu\nu}|^2
\frac{\ln\lambda_\mu-\ln\lambda_\nu}{\lambda_\mu-\lambda_\nu}
+O(\|Y\|^3),
\label{eq:supp_entropy_weighting}
\end{equation}
with the coincident-eigenvalue limit equal to $1/\lambda_\mu$.  Since $Y$ contains
only blocks connecting different subsystem charge sectors, the sum couples
eigenvectors from different sectors.  In the same basis,
$\CA=\sum_{\mu,\nu}|Y_{\mu\nu}|^2$.  The two diagnostics therefore contain the same
off-block matrix elements but weight them differently: every term has unit
weight in $\CA$, whereas its contribution to $\DS$ depends on the spectrum of
$\rhoAsym$.  Equation~\eqref{eq:supp_relative_entropy} is exact at all times.
The entanglement-asymmetry curves are calculated from the exact entropy
difference; the quadratic expansion in Eq.~\eqref{eq:supp_entropy_weighting}
is used to diagnose the channel weights.

For the block-resolved analysis at $L_A=5$, we label a charge sector by the
number of up spins $n=0,\ldots,5$, equivalently $Q_A=n-L_A/2$.  The directed
Frobenius contribution is
\begin{equation}
W_{nn'}=\left\|P_n\rho_A P_{n'}\right\|_F^2,
\qquad C_A=\sum_{n\ne n'}W_{nn'} .
\label{eq:supp_block_weight}
\end{equation}
Writing $|n,\alpha\rangle$ for an eigenvector of the $n$ block of
$\rhoAsym$, the corresponding quadratic entropy contribution is
\begin{equation}
K_{nn'}=\frac12\sum_{\alpha,\beta}
\left|\langle n,\alpha|Y|n',\beta\rangle\right|^2
\frac{\ln\lambda_{n\alpha}-\ln\lambda_{n'\beta}}
{\lambda_{n\alpha}-\lambda_{n'\beta}} .
\label{eq:supp_block_entropy_weight}
\end{equation}
Thus $\sum_{n\ne n'}K_{nn'}$ gives the quadratic approximation in
Eq.~\eqref{eq:supp_entropy_weighting}.

We compare the two diagnostics on identical trajectories: both use the same
field strength $\epsilon=0.6$, phase $\phi=0.37$, initial angles,
$0\leq Jt\leq30$ with 181 time points, position averaging, and
sustained-crossing rule.  We
take
$(N,L_A)=(12,3),(14,3),(16,4),(18,5),$ and $(20,5)$.  The detuned wave vectors
are fixed at $q/\pi=1.028$, as in Fig.~3.  We denote the sustained crossing fractions of the
raw $\CA(t)$ and $\DS(t)$ curves by $f_C$ and $f_S$, respectively.  Table~\ref{tab:coherence_crossings}
shows that the uniform field crosses in neither diagnostic, the generic and
staggered fields cross in both, and the detuned-staggered field has $f_C>0$ but
$f_S=0$ at every displayed size.

The block-resolved calculation was performed for the $N=20$, $L_A=5$ generic
and detuned-staggered fields and averaged over the same 16 contiguous
subsystems.  Direct reconstruction agrees with the saved observables within
the numerical tolerance used for crossings, and the quadratic result remains
stable as the eigenvalue cutoff is varied from $10^{-12}$ to $10^{-8}$.  The
main approximation is therefore the quadratic expansion itself.

\begin{table}[b]
\caption{Sustained crossing fractions $f_C/f_S$ for the raw Frobenius
charge-sector coherence and entanglement asymmetry.  Each fraction is evaluated
over the 15 pairs with unequal initial values, and a crossing must persist for at least
one unit of $Jt$.}
\label{tab:coherence_crossings}
\begin{ruledtabular}
\begin{tabular}{ccccc}
$N$ & uniform & generic & detuned-staggered & staggered \\
\hline
12 & $0/0$ & $1/1$ & $(10/15)/0$ & $1/1$ \\
14 & $0/0$ & $1/1$ & $(7/15)/0$ & $1/1$ \\
16 & $0/0$ & $1/1$ & $(11/15)/0$ & $1/1$ \\
18 & $0/0$ & $1/1$ & $(9/15)/0$ & $1/1$ \\
20 & $0/0$ & $1/1$ & $(6/15)/0$ & $1/1$ \\
\end{tabular}
\end{ruledtabular}
\end{table}

\begin{figure}[t]
\centering
\includegraphics[width=0.70\textwidth]{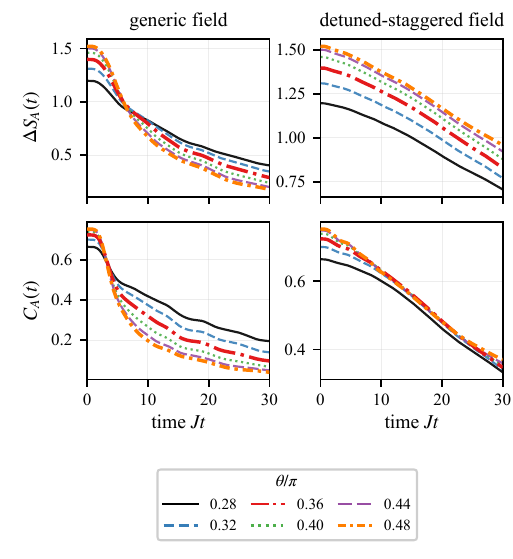}
\caption{Representative raw curves underlying the inversion-area comparison in
main-text Fig.~4.  The upper row shows $\DS(t)$ and the lower row shows the
unnormalized $\CA(t)$ for the $N=20$, $L_A=5$ generic and detuned-staggered
fields at $\epsilon=0.6$.  Colors denote the six initial angles $\theta/\pi$;
the thicker curves highlight the pair $0.36,0.48$ used in Fig.~\ref{fig:block_mechanism}.
The generic field crosses in both diagnostics; the
detuned-staggered field crosses in $\CA(t)$ but not in $\DS(t)$.}
\label{fig:coherence_mechanism}
\end{figure}

To compare inversion strengths, for either $X=\CA$ or $X=\DS$ we define
\begin{equation}
R_X^*=\max_{a,b:\,X_a(0)>X_b(0)}
\frac{1}{T}\int_0^T dt\,\bigl[X_b(t)-X_a(t)\bigr]_+,
\label{eq:supp_inversion_area}
\end{equation}
where $[x]_+=\max(x,0)$ and $T=30$.  Thus $R_X^*$ is the largest positive
inversion area among all pairs satisfying $X_a(0)>X_b(0)$, normalized by the total time.
Figure~\ref{fig:coherence_mechanism} displays the unnormalized $\CA(t)$
curves used to evaluate $f_C$.  In the generic field, coherence reverses its
ordering before the entanglement asymmetry, and both crossings persist.  In
the detuned-staggered field, only the coherence ordering reverses within the
simulated window.

\begin{figure}[t]
\centering
\includegraphics[width=0.82\textwidth]{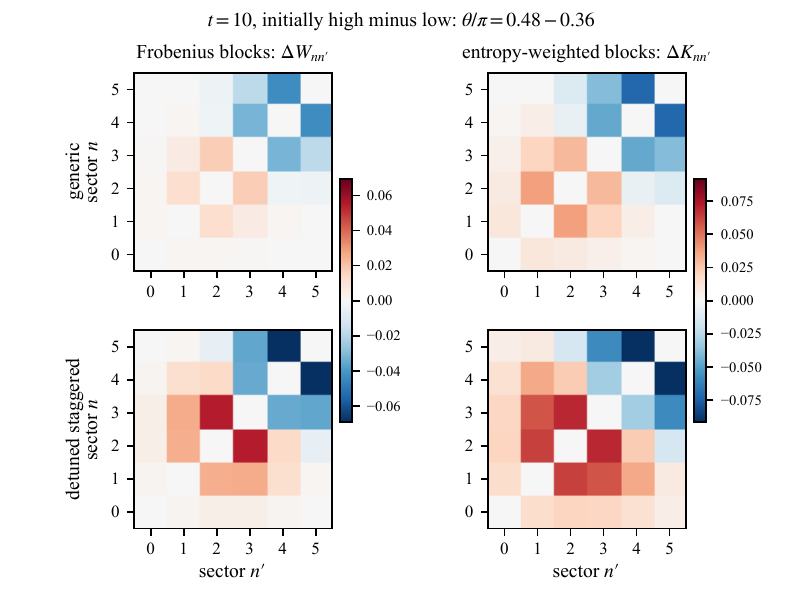}
\caption{Block-resolved ordering difference for the $N=20$, $L_A=5$ pair
$\theta/\pi=0.48$ (initially high) and $0.36$ (initially low) at $t=10$.
The upper and lower rows show the generic and detuned-staggered fields.  The
left column gives the directed Frobenius contributions $\Delta W_{nn'}$ and the
right column their entropy-weighted quadratic counterparts $\Delta K_{nn'}$.
Here $\Delta W_{nn'}=W_{nn'}^{\rm high}-W_{nn'}^{\rm low}$ and
$\Delta K_{nn'}=K_{nn'}^{\rm high}-K_{nn'}^{\rm low}$.
Blue entries favor an ordering inversion and red entries preserve the initial
ordering.  High-charge edge blocks are negative in both fields, while positive
central-sector contributions remain visibly stronger for the detuned-staggered
field.}
\label{fig:block_mechanism}
\end{figure}

Figure~\ref{fig:block_mechanism} resolves the charge-block competition for
the pair $\theta/\pi=0.36,0.48$.  In the detuned-staggered field, this pair
reverses its total coherence ordering but not its entanglement-asymmetry
ordering.  We use the same pair in the generic field for comparison.
Let $\delta X=X_{\rm high}-X_{\rm low}$ and
$K=\sum_{n\ne n'}K_{nn'}$.  At $t=10$, the exact and quadratic differences are
\begin{equation}
\begin{array}{c|rrr}
 & \delta C_A & \delta(\DS) & \delta K \\
\hline
\text{generic} & -0.124 & -0.121 & -0.119 \\
\text{detuned-staggered} & -0.00465 & +0.113 & +0.267
\end{array}
\end{equation}
Both fields invert the total coherence ordering, but only the generic field
inverts the entropy ordering.  Figure~\ref{fig:block_mechanism} traces the
difference to where the coherence resides.  High-charge edge blocks favor
inversion in both cases.  They outweigh the central blocks in the generic
field, whereas stronger central and intermediate contributions retain the
initial entropy ordering in the detuned-staggered field.  The entropic outcome
therefore depends on how coherence is distributed among charge blocks, not on
its total Frobenius weight alone.

For this representative pair, the detuned-staggered coherence inversion
persists from $t=8.5$ to $22$ on the sampled grid, while its exact entanglement
asymmetry remains ordered throughout $0\le t\le30$.  In the generic field,
the exact entropy crossing begins at $t=6$, compared with $t=6.5$ for the
quadratic diagnostic.  Across the full time grid, the quadratic and exact
entropy orderings agree almost everywhere in the generic field and everywhere
in the detuned-staggered field.

Across the six states and the full time grid, the median state-level relative
error of the quadratic expansion is $7.10\%$ for the generic field and
$11.83\%$ for the detuned-staggered field, with a maximum of $26.95\%$.
Errors in a difference between two states can be larger: at $t=10$, the
quadratic detuned-staggered result overestimates the exact entropy difference
by more than a factor of two, although it gives the correct sign.  We therefore
use the expansion to identify the competing block contributions, while all
crossing statements follow from the exact entropy curves.

The charge populations also enter this weighting through the eigenvalues of
$\rhoAsym$.  For the displayed pair at $t=10$, the population-difference
vectors $p_n^{\rm high}-p_n^{\rm low}$, with $p_n=\mathrm{Tr}(P_n\rho_A)$,
have cosine similarity $0.99213$ between the two fields.  Their directions
are close, but this comparison does not equate their magnitudes or the spectra
within individual charge blocks.  The block analysis retains both the
coherences and these eigenvalue-dependent weights.

\FloatBarrier
\bibliography{references}